\documentstyle[12pt]{article}
\begin{document}

\null\vskip-1cm
\rightline{hep-ph/9608346}
\rightline{August 1996}

\vspace{1cm}

\begin{center}

{\large\bf MULTIPLICITY DISTRIBUTIONS IN} \\
{\large\bf STRONG INTERACTIONS: A GENERALIZED} \\
{\large\bf NEGATIVE BINOMIAL MODEL}

\vspace{1.cm}

{\sc S. Hegyi}\footnote{E-mail: hegyi@rmki.kfki.hu}

\vspace{.3cm}

{\normalsize
	KFKI Research Institute for \\
	Particle and Nuclear Physics of the \\
	Hungarian Academy of Sciences, \\
	H--1525 Budapest 114, P.O. Box 49. \\
	Hungary
}

\end{center}

\vspace{.5cm}

\begin{abstract}
   A three-parameter discrete distribution is developed to
   describe the multiplicity distributions observed in total-
   and limited phase space volumes in different collision
   processes. The probability law is obtained by the Poisson
   transform of the KNO scaling function derived in
   Polyakov's similarity hypothesis for strong interactions
   as well as in perturbative QCD,
   $\psi(z)\propto z^\alpha\exp(-z^\mu)$.
   Various characteristics of the newly proposed distribution
   are investigated e.g. its generating function, factorial moments,
   factorial cumulants. Several limiting and special cases
   are discussed. A comparison is made to the multiplicity data
   available in $e^+e^-$ annihilations at the $Z^0$ peak.
\end{abstract}

\newpage
\pagestyle{plain}
\noindent
{\LARGE\bf 1. Introduction}

\vspace{.5cm}
\noindent
Over the past years much attention has been focused on the description
of multiplicity distributions (MDs) for hard processes
in perturbative quantum
chromodynamics (pQCD), see ref.~[1] for recent reviews. Already in the
lowest double logarithmic approximation (DLA) it was demonstrated that
QCD provides a natural explanation of the KNO scaling law
$$
	P_n=\frac{1}{\langle n\rangle}\,
	\psi\left(\frac{n}{\langle n\rangle}\right)
	\eqno(1.1)
$$
of the MDs~[2] but the width of the scaling function
$\psi(z=n/\bar{n})$ was overestimated. To improve on the agreement with
experiments higher-order perturbative corrections should be taken into
account. Mathematically this manifests itself in the appearance of a new
expansion parameter $\gamma q$,
the product of the QCD multiplicity anomalous
dimension $\gamma\propto\sqrt{\alpha_s}$
and the rank $q$ of the moments of $P_n$. For
example, the influence of recoil effects~[3] on multiplicity fluctuations
is seen most clearly when one considers the normalized factorial moments
$$
	F_q=\frac{\langle n(n-1)\ldots(n-q+1)\rangle}
	{\langle n\rangle^q}.\eqno(1.2)
$$
In the next-to-next-to-leading log approximation (NNLA) the $F_q$ are
shown~[4] to be dominated for large ranks~$q$ by the
\hbox{$\Gamma$-function} of the rescaled rank $q/\mu$,
$$
	F_q\propto\Gamma(\mbox{$\frac{3}{2}$}+q/\mu)
        \, D^{-q}\eqno(1.3)
$$
where $\mu=(1-\gamma)^{-1}$ and $D$ is a scale parameter depending on $\mu$.
Since $\gamma\simeq0.4$ at the $Z^0$ peak,
the strength of higher-order multiplicity fluctuations are
drastically reduced by recoil effects and therefore the tail of the
KNO function gets modified. Instead of the exponential fall-off
predicted in the DLA, a faster than exponential decay law emerges for
large~$z$~[4]
$$
	\psi(z)\propto z^\alpha\exp(-[Dz]^\mu)\eqno(1.4)
$$
with $\alpha=3\mu/2-1$ and $\mu\simeq5/3$.
This type of behaviour reduces the width of
$\psi(z)$ overestimated by the early DLA calculations.

Another manifestation of higher-order pQCD effects on the MDs
is best seen in the $q$-dependence of the ratio $H_q=K_q/F_q$ invented
by Dremin~[5] where $K_q$ denotes the normalized factorial cumulant
moments,
$$
	K_q=F_q-\sum_{i=1}^{q-1}{q-1\choose i}K_{q-i}F_i\,.
	\eqno(1.5)
$$
In the DLA $H_q$ decreases monotonously as $q^{-2}$ but the inclusion
of higher order perturbative corrections yields quite nontrivial
$q$-dependence. In the next-to-leading log approximation (MLLA)
$H_q$ acquires a negative minimum at $q\simeq5$ and approaches for
large ranks~$q$ the abscissa from below. In the NNLA the negative minimum
at $q\simeq5$ is followed by sign-changing
oscillations of the moment ratios~[6]. Quite recently this peculiar
behaviour has been experimentally confirmed in $e^+e^-$ annihilations~[7].

In a historical context it is worth mentioning that some of the above
predictions of pQCD were put forward already in 1970 by Polyakov
formulating a similarity hypothesis for strong interactions
in $e^+e^-$ annihilations~[8].
He obtained asymptotic KNO scaling behaviour for the MDs
(two years earlier than the original KNO-paper) with a scaling function
that behaves according to
$$
	\psi(z)\Rightarrow\left\{\begin{array}{ll}
	\exp(-z^{1/(1-2\delta)}) & \mbox{as $z\to\infty$} \\
	0		         & \mbox{as $z\to0$}
	\end{array}\right.\eqno(1.6)
$$
where $0<\delta<1/2$ and the small-$z$ behaviour of the scaling
function was found, with certain assumptions, to be a monomial~[8].
Clearly, Polyakov arrived in his model at the same law for
multiplicity fluctuations as the pQCD result~(1.4).
A KNO scaling function of the form of Eq.~(1.4)
was studied in more or less detail by others as well~[9]
most notably by Krasznovszky and Wagner.
We call attention also to the remarks made by
Koba, Nielsen and Olesen in connection with the moment problem~[2]
to which we shall return later.

The goal of the present paper is the development of a discrete
distribution well suited to describe available multiplicity data
when the popular negative binomial parametrization fails.
Our guiding principle is the incorporation of the pQCD-based
characteristics
{\it i)\/}~factorial moments $F_q$ dominated by the
\hbox{$\Gamma$-function}
of the rescaled rank $q/\mu$,
{\it ii)\/}~factorial cumulant-to-moment ratios $H_q$ displaying
nontrivial sign-changing oscillations, and
{\it iii)\/}~distributions $P_n$
possessing squeezed high-multiplicity
tail in accordance with Eq.~(1.4).
We shall see that a simple generalization
of the negative binomial law with one additional parameter
satisfies all these requirements.

\newpage
\noindent
{\LARGE\bf 2. The scaling function in pQCD}

\vspace{.5cm}
\noindent
Let us first consider in  more detail the
modified KNO scaling function~(1.4) obtained in pQCD with
fixed coupling approximation~[4].
For later convenience we use a somewhat different choice of the
parameters. Introducing the shape parameter
\hbox{$k=(\alpha+1)/\mu$} and
denoting the scale parameter by $\lambda$ we get
$$
	\psi(z)={\cal N}\,z^{\mu k-1}
	\exp\left(-[\lambda z]^{\mu}\right)
	\eqno(2.1)
$$
where ${\cal N}$ is a normalization constant. The scaling function
$\psi(z)$ should satisfy the two normalization conditions
$$
	\int_0^\infty \psi(z)\,dz=\int_0^\infty z\,\psi(z)\,dz=1.
	\eqno(2.2)
$$
The first condition determines the constant ${\cal N}$.
It can be obtained by the use of the integral~[10]
$$
	\int_0^\infty
	z^{\mu k-1}\exp\left(-[\lambda z]^{\mu}\right)dz=
	\frac{\Gamma(k)}{\mu\lambda^{\mu k}}\eqno(2.3)
$$
providing the reciprocal of ${\cal N}$. Thus the complete form of
the scaling function (at the moment without the second normalization
constraint) is
$$
	\psi(z)=\frac{\mu}{\Gamma(k)}\,\lambda^{\mu k}
	z^{\mu k-1}\exp\left(-[\lambda z]^{\mu}\right).\eqno(2.4)
$$
Eq.~(2.4) is known in the mathematical literature
as the generalized gamma distribution~[11]. Obviously, the $\mu=1$
special case yields the ordinary gamma distribution which corresponds
in pQCD to the KNO scaling function at asymptotically high energies
($\gamma\propto\sqrt{\alpha_s}\to0$).

The probability density function (2.4) has a long history~[12].
Its first appearance in 1925 is due to L. Amoroso who analyzed the
distribution of economic income. The next application is
related to the grinding of~materials:
in~1933 Rosin, Rammler and Sperling arrived at
Eq.~(2.4) as the size distribution of grains
produced by comminution.
Nevertheless in the literature of particle size distributions
the formula is named after Nukiyama and Tanasawa who rediscovered it
in 1939 studying drop size distributions in sprays.
Widespread use of Eq.~(2.4) by mathematicians started only in 1962
by the paper of Stacy. He introduced it as the generalized
gamma distribution and since then this name is the most frequently
used one. Some of the recent applications include
polymer physics~[13] and the theory of multiplicative processes~[14].

The moments of the scaling function (2.4) are readily obtained from the
integral given by Eq.~(2.3) which yields
$$
	\langle z^q\rangle=\int_0^\infty z^q\,\psi(z)\,dz=
	\frac{\Gamma(k+q/\mu)}{\Gamma(k)}\frac{1}{\lambda^q}\,.
	\eqno(2.5)
$$
Eq.~(2.5) with $q=1$ and
the second normalization condition in Eq.~(2.2) restrict
the scale parameter to
$$
	\lambda=\frac{\Gamma(k+1/\mu)}{\Gamma(k)}\eqno(2.6)
$$
which completes the analytic form of the scaling function.
Let us now consider the cumulative distribution corresponding
to the probability density~(2.4). It is given by
$$
	\varphi(x)=\int_0^x\psi(z)\,dz=
	\frac{\gamma\left(k,[\lambda x]^\mu\right)}
	{\Gamma(k)}\eqno(2.7)
$$
where $\gamma(\cdot)$ denotes the incomplete gamma function. Eq.~(2.7)
is particularly useful in connection with KNO-G scaling [15,16] i.e. if
the discrete multiplicity distributions $P_n$ are approximated by the
continuous scaling function~$\psi(z)$,
$$
	P_n=\int_{z_n}^{z_{n+1}}\psi(z)\,dz=
	\varphi(z_{n+1})-\varphi(z_n).\eqno(2.8)
$$
Although KNO-G scaling is becoming increasingly popular in the
description of MDs, with $\psi(z)$ chosen to be a shifted lognormal
density, we will follow a different approach to
construct $P_n$ from the scaling function.

Perhaps the most
advantageous property of the generalized gamma distribution is
that $\psi(z;k,\lambda,\mu)$ can be used to specify a variety of well known
probability laws. Some examples are cited in Table~1 with the corresponding
set of parameters. It is seen that the exponent $\mu$ can take
negative values as well in which case the normalization constant in
Eq.~(2.4) involves~$|\mu|$. Moreover, for negative $\mu$ the moments
$\langle z^q\rangle$ of the distribution are finite only for
ranks~$q$ satisfying $q/\mu>-k$. A good example is the completely
asymmetric L\'evy law of index $\alpha=1/2$ having no
finite, positive-rank moments. Finally let us
call special attention to the last two
rows of Table~1 that display the limit distributions of
$\psi(z;k,\lambda,\mu)$ involving for $\mu\to0$ the lognormal law.

\newpage
\noindent
{\LARGE\bf 3. MDs at preasymptotic energies}

\vspace{.5cm}\noindent
Comparing the KNO function~(2.4) to the pQCD result quoted in
Section~1 we see that the more accurate account of
recoil effects in~[4] yields for
$\psi(z)$ a generalized gamma distribution with fixed shape parameter
$k=3/2$. Furthermore, the factorial moments given by Eq.~(1.3) are the
same as the ordinary moments~(2.5) of
$\psi(z;\frac{3}{2},\lambda,\mu)$ which is due to the identification of
$\langle z^q\rangle$ and $F_q$ in ref.~[4]. Their equivalence holds
valid for the $\mu=1$ asymptotic energy scale as well as
for preasymptotic energies characterized by $\mu>1$
if $P_n$ is defined by the Poisson transform of $\psi(z)$,
$$
   	P_n=\int_0^\infty \psi(z)\,\frac{(\bar{n}z)^n}{n!}
   	e^{-\bar{n}z}\,dz.\eqno(3.1)
$$
For Eq.~(3.1) the asymptotic scaling limit of the MDs is~[17]
$$
	\lim_{
	n\rightarrow\infty,\,\langle n\rangle\rightarrow\infty
	\atop n/\langle n\rangle\mbox{\ \tiny fixed}}
	P_n=\frac{1}{\langle n\rangle}\,
	\psi\left(\frac{n}{\langle n\rangle}\right).
	\eqno(3.2)
$$
Thus the knowledge of the shape of $P_n$ at preasymptotic energies
enables one to guess the asymptotic shape of $\psi(z)$.
We mention that the scaling limit~(3.2) naturally arises
in Polyakov's model~[18] suggesting the use of~(3.1) instead of~(2.8)
to construct $P_n$ from~$\psi(z)$.

The probability generating function of $P_n$ defined by the Poisson
transform~(3.1) reads as follows:
$$
	{\cal G}(u)=\sum_{n=0}^\infty(1-u)^n P_n=
	\int_0^\infty \psi(z)\,e^{-u\bar{n}z}dz,
	\eqno(3.3)
$$
i.e. ${\cal G}(u)$ is given
by the Laplace transform of $\psi(z)$~[17]. Unfortunately the
Laplace transform of Eq.~(2.4) can be expressed in terms of
special functions only for a few specific values of the parameter~$\mu$.
To go further we have to make use of Fox's generalized hypergeometric
function, ${\sf H}(x)$. The reader unfamiliar with the
theory of Fox functions
can find a compilation of the necessary formulae in the Appendix. The
KNO scaling function~(2.4) is expressed in terms of ${\sf H}(x)$ as
the following ${\sf H}$-function distribution:
$$
   \psi(z)=\frac{\lambda}{\Gamma(k)}\;{\sf H}^{1,0}_{0,1}
   \left[\,\lambda z\left|
      \begin{array}{c}
         -    		     	\\
	 (k-1/\mu,\; 1/\mu)
      \end{array}
   \right]\right..\eqno(3.4)
$$
The above form is obtained from Eq.~(A.7) using the identity~(A.4) and
the integral~(2.3) to ensure proper normalization, see also~[22].
The probability generating function of $P_n$ defined by the Poisson
transform of Eq.~(3.4) can be easily evaluated with the help of~(A.5),
the Laplace transform of ${\sf H}(x)$. It yields
$$
   {\cal G}(u)=
   \frac{1}{\Gamma(k)}\;{\sf H}^{1,1}_{1,1}
   \left[
      \frac{\lambda}{u\bar{n}}\left|
      \begin{array}{c}
         (1,\; 1)     	\\
	 (k,\; 1/\mu)
      \end{array}
   \right],\right.\;0<\mu<1\eqno(3.5)
$$
and
$$
   {\cal G}(u)=
   \frac{1}{\Gamma(k)}\;{\sf H}^{1,1}_{1,1}
   \left[
      \frac{u\bar{n}}{\lambda}\left|
      \begin{array}{c}
         (1-k,\; 1/\mu)     	\\
	 (0,\; 1)
      \end{array}
   \right],\right.\;\mu>1\eqno(3.6)
$$
where $\lambda$ is given by Eq.~(2.6). The necessity of two separate
expressions for ${\cal G}(u)$ follows from the existence conditions of
${\sf H}(x)$ discussed in the Appendix. The $\mu=1$ case is of course
the generating function of the negative binomial distribution,
${\cal G}(u)=(1+u\bar{n}/k)^{-k}$. The $\mu<0$ case will be studied
elsewhere.

The characteristic function
$\phi(t)$ of the generalized gamma density~(2.4)
is usually written as an infinite sum~[21]. In terms of the Fox function
$\phi(t)$ can be expressed in a much simpler way by changing the variable
in ${\cal G}(u)$ given by~Eqs.~(3.5-6),
$$
	\phi(t)=\int_0^\infty\psi(z)e^{itz}dz=
	{\cal G}\left(u\bar{n}\Rightarrow-it\right)
	\eqno(3.7)
$$
with, in general, unconstrained scale parameter~$\lambda$.
For $\mu>1$ $\phi(t)$ was obtained earlier in ref.~[22]. The probability
generating function provides also the Poisson transform of $\psi(z)$
for $n=0$ through \hbox{$P_0={\cal G}(1)$.}

Let us now consider the Poisson transform of~(3.4) for arbitrary $n$.
By simple algebra, using identities (A.2-4) and the Laplace
transform~(A.5) of ${\sf H}(x)$ one arrives at the
following discrete ${\sf H}$-function distributions:
$$
   P_n=
   \frac{1}{n!\,\Gamma(k)}\;{\sf H}^{1,1}_{1,1}
   \left[
      \,\frac{\lambda}{\bar{n}}\left|
      \begin{array}{c}
         (1-n,\; 1)     	\\
	 (k,\; 1/\mu)
      \end{array}
   \right],\right.\;0<\mu<1\eqno(3.8)
$$
and
$$
   P_n=
   \frac{1}{n!\,\Gamma(k)}\;{\sf H}^{1,1}_{1,1}
   \left[
      \,\frac{\bar{n}}{\lambda}\left|
      \begin{array}{c}
         (1-k,\; 1/\mu)     	\\
	 (n,\; 1)
      \end{array}
   \right],\right.\;\mu>1.\eqno(3.9)
$$
The $\mu=1$ case yields the Poisson transform of the ordinary gamma
distribution, i.e. the negative binomial law.
The splitted parameter space for~$\mu$ reflects an
important difference between the two expressions for~$P_n$. In case of
Eq.~(3.8) $P_n$ is infinitely divisible, just as the negative
binomial for $\mu=1$, whereas the $\mu>1$ case given by Eq.~(3.9)
violates this feature. Due to the preservation of infinite
divisibility under Poisson transforms the same distinction
between the two $\mu$-domains holds for the
generalized gamma distribution~[23].

The factorial moments of $P_n$ can be determined through the equivalence
of $\langle z^q\rangle$ and $F_q$ for Poisson transforms~[17].
Plugging into Eq.~(2.5) with scale parameter $\lambda$ restricted
by~(2.6) we get
$$
   	F_q=\frac{\Gamma(k+q/\mu)}{\Gamma^q(k+1/\mu)}\,
   	\Gamma^{q-1}(k).\eqno(3.10)
$$
As intended, the factorial moments are dominated by the $\Gamma$-function
of the rescaled rank $q/\mu$ for large~$q$.
With the help of Eq.~(1.5) we have calculated the moment ratios
$H_q=K_q/F_q$ for shape parameter $k=3/2$.
The behaviour of $\log |H_q|$ over the $\mu$-$q$ plane is shown in
Fig.~1a. The peculiar $q$-dependence for $\mu>1$ is due to sign-changing
oscillations of $H_q$. Observe that the pattern of oscillations is
nontrivial, i.e. not alternating as $q$ takes even/odd values,
see Fig.~1b. Qualitatively similar sign-changing oscillations of $H_q$
occur for a different choice of the shape parameter~$k$.

\vspace{.9cm}\noindent
{\LARGE\bf 4. Comparison to $e^+e^-$ data}

\vspace{.6cm}
\noindent
One of the advantages of $P_n$ given by Eqs.~(3.8-9) is its generality:
the Poisson transform of many well known probability laws (e.g.
of those cited in \hbox{Table~1)} is a special case.
The negative binomial distribution ($\mu=1$)
not rarely fails to give a reasonable description of the
MDs. For example, in $e^+e^-$ annihilations the full
phase-space MDs at the $Z^0$ peak show significant deviation from a
negative binomial shape as reflected by the
\hbox{$\chi^2/\mbox{d.o.f.}=66/23$} and 68/24 values reported by the
Delphi and SLD collaborations~[24,7]. Moreover, the sign-changing
oscillations displayed by the SLD data for $H_q$~[7] can not be
reproduced by a (possibly truncated) negative binomial.
It is natural to ask how the $\mu\neq1$ special
cases of Eqs.~(3.8-9) can describe the observed features.

To answer this question we carried out fits to the $e^+e^-$ multiplicity
data at the $Z^0$ peak~[7,24]. In the fitting procedure numerical
evaluation of the integral~(3.1) is implemented using \hbox{96-point}
gaussian quadrature. The scaling function~(2.4) is log-linear so that
$y=\ln z$ can be written as \hbox{$y=w/\mu-\ln\lambda$} where
$w$ has probability density
$
	f(w)=\exp(kw-e^w)/\Gamma(k).
$
Making the parameter transformations $p=k^{-1/2}$, $\sigma=p/\mu$
and $\alpha=\ln\lambda$, further, reflecting the model about the origin
$p=0$ to negative~$p$ the scaling function~(2.4) takes the form
$$
	\psi(z)=\frac{|p|}{\Gamma(p^{-2})\,\sigma z}\,
	\exp\left[p^{-2}w-e^w\right]\qquad\mbox{if $p\neq0$}
	\eqno(4.1)
$$
with
$
	w=p\,(\ln z+\alpha)/\sigma.
$
The reparametrization allows the lognormal law to be mapped to the
origin,
$$
	\psi(z)=\frac{1}{\sqrt{2\pi}\,\sigma z}\,
	\exp\left[-\frac{1}{2}\frac{(\ln z+\alpha)^2}
	{\sigma^2}\right]\quad\mbox{if $p=0$}.\eqno(4.2)
$$
Above, the location parameter $\alpha$ is restricted by the second
normalization condition in Eq.~(2.2) to
$$
	\alpha=\left\{\begin{array}{ll}
  \ln\Gamma(p^{-2}+\sigma/p)-\ln\Gamma(p^{-2}) &\qquad\mbox{if $p\neq0$} \\
  \sigma^2/2				       &\qquad\mbox{if $p=0$}
	\end{array}\right..\eqno(4.3)
$$
The distributions represented by Eqs.~(4.1-2) are stochastically
continuous in the parameters and include besides the lognormal ($p=0$)
e.g. the gamma ($p=\sigma$), exponential ($p=\sigma=1$)
and Weibull ($p=1$) densities. For
further details on the above transformation of generalized gamma
variates the reader is referred to~[25].

In the numerical evaluation of the integral~(3.1) we have used
\hbox{Eqs.~(4.1-2)}
for $\psi(z)$. There is a strong reason of replacing the
parametrization~(2.4) with those of (4.1-2) in the fitting procedure.
The original fits using the form~(2.4)
were able to reduce significantly
the $\chi^2$ corresponding to the $\mu=1$ negative binomial case but they
resulted unexpectedly small values of~$\mu$. For example, fitting the
Delphi data yields $\chi^2/\mbox{d.o.f.}=33/23$, $k\simeq1000$ and
$\mu\simeq0.1$ indicating that the observed $P_n$ is a Poisson
transformed lognormal distribution. Performing the fits using
(4.1-2) reinforced our finding for each data set. This can
be seen from Table~2 where the outcome of the fits are collected.
In Fig.~2a the best-fit theoretical $P_n$ is displayed for the SLD data.

The above result seems to be in sharp contradiction with the SLD data for
the moment ratios $H_q$~[7]. Sign-changing oscillations can occur in our
model only for $\mu>1$, if $P_n$ is not infinitely divisible. Fitting the
$H_q$ data with the help of Eqs.~(3.10) and (1.5) produced unacceptable
$\chi^2$. To allow the possible influence of truncation effects~[26]
we recalculated the factorial moments $F_q$ in terms of $P_n$ using
$n_{min}=6$ and $n_{max}=50$ in accordance with the SLD data~[24].
The resulting behaviour of $H_q$ is shown in Fig.~2b, the agreement
between the theoretical curve and the data points is excellent. The
corresponding $\chi^2$ and the values of the best-fit parameters are cited
in the last row of Table~2. Clearly, the obtained parameters are
very similar to those found in the previous fits.
Although there is a small but significant
deviation from $p=0$ we can safely conclude that the multiplicity data
in $e^+e^-$ annihilations at the $Z^0$ peak favour
a lognormally shaped  KNO function $\psi(z)$
in the asymptotic limit given by Eq.~(3.2).

\vspace{.9cm}\noindent
{\LARGE\bf 5. Summary and conclusions}

\vspace{.5cm}
\noindent
We have developed a three-parameter discrete distribution to analyse the
experimental data for MDs in different collision processes. It is obtained
by the Poisson transform of the generalized gamma density~(2.4) and thus
it provides a generalization of the popular negative binomial distribution.
The analytic form of $P_n$ is expressed in terms of ${\sf H}$-functions.
It is given by Eq.~(3.8) for infinitely divisible $P_n$ and by Eq.~(3.9)
if $P_n$ does not obey this feature. The $\mu=1$ marginal
case between the two families of
distributions is the negative binomial law. Violation
of infinite divisibility for Eq.~(3.9) allows nontrivial sign-changing
oscillations of the factorial cumulant-to-moment ratios $H_q$. For
$\mu>1$ we thus have the ability to reproduce this peculiar $q$-dependence
of $H_q$ data without truncation. Since the factorial moments~(3.10) are
dominated by the $\Gamma$-function of the rescaled rank $q/\mu$ for
large~$q$ we can also reproduce possible enhancement ($\mu<1$) and
suppression ($\mu>1$) of multiplicity fluctuations with respect to the
negative binomial behaviour.

Fitting the newly developed distribution to the $e^+e^-$ multiplicity
data at the $Z^0$ peak we have found $\mu<1$ departure from the
negative binomial law. In fact
the observed MDs are in agreement with the Poisson transform of the
$\mu\to0$ limit distribution of~(2.4) which is the lognormal density.
Even the sign-changing oscillations of the SLD data for $H_q$ can be
fully reproduced after taking into account truncation effects.
It is important to emphasize that the $\mu\to0$ limit manifests itself
through the factorial moments of $P_n$ being equivalent to the ordinary
moments of the asymptotic $\psi(z)$. At the $Z^0$ peak the KNO function
exhibits suppressed high-multiplicity tail
in accordance with the pQCD results quoted earlier.
We disagree with ref.~[16] stating that $\psi(z)$
is lognormally shaped at current energies, which became a common
belief in recent years. Fitting the scaling
function~(2.4) to $\psi(z)$ observed by experiments displays clear
deviation from lognormality yielding $k\simeq5$ and
$\mu\simeq1.6$ best-fit parameters.
The marked difference between the experimental and asymptotic scaling
functions warns that the ordinary and factorial moments of $P_n$ can not
be identified at current energies and sign-changing oscillations caused
by pQCD effects are more likely to occur for the ordinary
cumulant-to-moment ratios.

The Poisson transformed lognormal shape of MDs calls attention to the
moment problem considered already in ref.~[2] for Eq.~(2.4).
According to our results the KNO function at the asymptotics~(3.2)
is not determined by its moments, a well known feature of the lognormal
law. There is another important aspect of lognormality
at asymptotic energies. It suggests that the limiting shape of $\psi(z)$
can be guessed on more elementary grounds than the perturbative
treatment of parton evolution equations.
The multiplicative nature of parton cascades in QCD
and the large number of steps of the cascade processes at very
high energies cause the central limit theorem to come into operation
determining uniquely the asymptotic shape of the scaling function.

\vspace{1.5cm}
\noindent
ACKNOWLEDGEMENTS

\vskip.5cm
\noindent
I am indebted to T. Cs\"org\H o and G. Jancs\'o for their useful
comments on the manuscript and to S. Krasznovszky for the
conversations on the subject. Further,
I would like to thank J. Zhou, SLD Collaboration, for communicating the
SLD multiplicity data. This work was supported by the Hungarian Science
Foundation under Grant No. OTKA-F4019/1992.

\newpage\null\noindent
{\LARGE\bf Appendix}

\vspace{.5cm}
\noindent
Here we give a brief summary of the basic properties of Fox's
generalized hypergeometric function, ${\sf H}(x)$.
It heavily relies on article~[19] and on the book~[20]
where the interested reader can find more details.

The ${\sf H}$-function of Fox is defined in terms of a
Mellin-Barnes type integral as follows:
$$
{\sf H^{m,n}_{p,\;q}}(x)=
{\sf H^{m,n}_{p,\;q}}\left[\,x\left|
      \begin{array}{c}
         (a_p,\; \alpha_p)     	\\
	 (b_q,\; \beta_q)
      \end{array}
   \right]\right.=\frac{1}{2\pi i}\int_{\cal L}h(s)\,x^s\,ds
   \eqno(\mbox{A.1})
$$
where $x\neq0$ and
$$
	x^s=\exp\left\{s\,[\,\ln |x|+i\arg(x)]\right\}
$$
in which $\arg(x)$ is not necessarily the principal value. Further,
$$
	h(s)=\frac
	{\prod_{j=1}^m\Gamma(b_j-\beta_js)\,
         \prod_{j=1}^n\Gamma(1-a_j+\alpha_js)}
	{\prod_{j=m+1}^q\Gamma(1-b_j+\beta_js)\,
	 \prod_{j=n+1}^p\Gamma(a_j-\alpha_js)}
$$
where $m,n,p,q$ are integers satisfying
$$
	0\leq n\leq p,\qquad1\leq m\leq q.
$$
The parameters $(a_1,\ldots,a_p)$ and $(b_1,\ldots,b_q)$ are complex,
whereas $(\alpha_1,\ldots,\alpha_p)$ and $(\beta_1,\ldots,\beta_q)$ are
positive numbers. An empty product is interpreted as unity. The
parameters are restricted by the condition that
	$\alpha_j(b_h+\nu)\neq\beta_h(a_j-1-\lambda)$
for $\nu,\lambda=0,1,\ldots$; $h=1,\ldots,m$; $j=1,\ldots,n$. The
contour ${\cal L}$ in the complex $s$ plane is such that the points
$s=(b_h+\nu)/\beta_h$ and $s=(a_j-1-\nu)/\alpha_j$ lie to the
right and left of ${\cal L}$ respectively while ${\cal L}$ extends from
$s=\infty-ik$ to $s=\infty+ik$ where $k$ is a constant with
$k>|{\rm Im\ } b_h|/\beta_h$.

The ${\sf H}$-function makes sense
only if the following two existence conditions are satisfied:

\noindent
{\it i)\ } $x\neq0$ and $\rho>0$ where
$$
	\rho=\sum_{j=1}^q\beta_j-\sum_{j=1}^p\alpha_j
$$

\noindent
{\it ii)\ } $\rho=0$ and $0<|x|<\theta^{-1}$ where
$$
	\theta=\prod_{j=1}^p\alpha_j^{\alpha_j}
	    \prod_{j=1}^q\beta_j^{-\beta_j}.
$$
Under these conditions ${\sf H}(x)$ is an analytic function for
$x\neq0$, in general multivalued, one-valued on the Riemann
surface of $\ln x$.

Elementary properties
of ${\sf H}(x)$ utilized in the body of the paper:
$$
	{\sf H^{m,n}_{p,\;q}}\left[\,x\left|
        \begin{array}{c}
	         (a_p,\; \alpha_p)     	\\
		 (b_q,\; \beta_q)
	\end{array}
	\right]\right.=
	{\sf H^{n,m}_{q,\;p}}\left[\,\frac{1}{x}\left|
        \begin{array}{c}
		 (1-b_q,\; \beta_q)	\\
	         (1-a_p,\; \alpha_p)
	\end{array}
	\right]\right.
	\eqno(\mbox{A.2})
$$
\
$$
        x^c\,{\sf H^{m,n}_{p,\;q}}\left[\,x\left|
        \begin{array}{c}
	         (a_p,\; \alpha_p)     	\\
		 (b_q,\; \beta_q)
	\end{array}
	\right]\right.=
	{\sf H^{m,n}_{p,\;q}}\left[\,x\left|
        \begin{array}{c}
	         (a_p+c\,\alpha_p,\; \alpha_p)     	\\
		 (b_q+c\,\beta_q,\; \beta_q)
	\end{array}
	\right]\right.
	\eqno(\mbox{A.3})
$$
\
$$
        {\sf H^{m,n}_{p,\;q}}\left[\,x^c\left|
        \begin{array}{c}
	         (a_p,\; \alpha_p)     	\\
		 (b_q,\; \beta_q)
	\end{array}
	\right]\right.=
	\frac{1}{c}\,{\sf H^{m,n}_{p,\;q}}\left[\,x\left|
        \begin{array}{c}
	         (a_p,\; \alpha_p/c)     	\\
		 (b_q,\; \beta_q/c)
	\end{array}
	\right]\right.
	\eqno(\mbox{A.4})
$$
where $c>0$.
The first identity, Eq.~(A.2), is an important one because it enables us
to transform ${\sf H}(x)$ with $\rho<0$ to ${\sf H}(1/x)$ with $\rho>0$
and  satisfying the existence condition {\it i)\ } given above.
The Laplace transform of ${\sf H}(x)$ reads as follows:
$$
	\int_0^\infty
	{\sf H^{m,n}_{p,\;q}}(\,cx)\,e^{-rx}dx=
	\frac{1}{c}\,
	{\sf H^{n+1,m}_{q,\;p+1}}
	\left[\,\frac{r}{c}\left|
        \begin{array}{c}
		 (1-b_q-\beta_q,\; \beta_q)	\\
		 (0,\;1),\;(1-a_p-\alpha_p,\; \alpha_p)
	\end{array}
	\right].\right.
	\eqno(\mbox{A.5})
$$
Special cases of the ${\sf H}$-function include such as
Meijer's ${\sf G}$-function, Bessel, Legendre, Whittaker,
Struve functions, the generalized hypergeometric function and several
others. Meijer's ${\sf G}$-function is obtained when the ${\sf H}$-function
parameters $(\alpha_1,\ldots,\alpha_p)$ and $(\beta_1,\ldots,\beta_q)$ are
unity. The generalized hypergeometric function $_pF_q$ is related
to ${\sf H}(x)$ through
$$
	_pF_q(a_p,b_q;x)=
	\frac{\prod_{j=1}^q\Gamma(b_j)}{\prod_{j=1}^p\Gamma(a_j)}\,
	{\sf H^{1,p}_{p,q+1}}\left[-x\left|
        \begin{array}{c}
		 (1-a_p,\; 1)	\\
		 (0,\;1),\;(1-b_q,\; 1)
	\end{array}
	\right].\right.
	\eqno(\mbox{A.6})
$$ From this relationship one can obtain numerous distributions of statistics
(associated e.g. with the Gauss- and confluent hypergeometric functions
$_2F_1$ and $_1F_1$) in terms of ${\sf H}(x)$.
Two further special cases of ${\sf H}(x)$ frequently
encountered in statistics are
$$
	x^\alpha\exp(-\lambda x^\beta)=
	\lambda^{-\alpha/\beta}\;
	{\sf H^{1,0}_{0,1}}\left[\lambda x^\beta\left|
        \begin{array}{c}
		  - 	\\
		 (\alpha/\beta,\; 1)
	\end{array}
	\right]\right.
	\eqno(\mbox{A.7})
$$
and
$$
	x^\alpha/(1+\lambda x^\beta)\;=\;
	\lambda^{-\alpha/\beta}\;
	{\sf H^{1,1}_{1,1}}\left[\lambda x^\beta\left|
        \begin{array}{c}
		 (\alpha/\beta,\; 1) 	\\
		 (\alpha/\beta,\; 1)
	\end{array}
	\right].\right.
	\eqno(\mbox{A.8})
$$
For a rich collection of particular cases of the ${\sf H}$-function, see
ref.~[20].

Let us now consider the continuous random variable
$x\in(0,\infty)$ whose probability density function is given by
$$
	f(x)={\cal N}\,
	{\sf H^{m,n}_{p,\;q}}\left[\,\lambda x\left|
      \begin{array}{c}
         (a_p,\; \alpha_p)     	\\
	 (b_q,\; \beta_q)
      \end{array}
   \right]\right.\eqno(\mbox{A.9})
$$
with scale parameter $\lambda>0$ and normalization condition
$
	\int_0^\infty f(x)\,dx=1.
$
The random variable $x$ is called a ${\sf H}$-function variate and the
probability law~(A.9) with $f(x)\geq0$
is the so-called ${\sf H}$-function distribution. Its parameters
$(a_p,\,\alpha_p)$ and $(b_q,\,\beta_q)$ should satisfy all the
restrictions given earlier in this Appendix. The characteristic function
of $f(x)$ can be obtained from Eq.~(3.7) and the Laplace transform (A.5):
$$
	\phi(t)=\frac{\cal N}{\lambda}
	{\sf H^{n+1,m}_{q,\;p+1}}
	\left[-\frac{it}{\lambda}\left|
        \begin{array}{c}
		 (1-b_q-\beta_q,\; \beta_q)	\\
		 (0,\;1),\;(1-a_p-\alpha_p,\; \alpha_p)
	\end{array}
	\right]\right..
	\eqno(\mbox{A.10})
$$
The $q$th moment of the ${\sf H}$-function distribution is
related to its $r=(q+1)$th Mellin transform through
$$
	\langle x^q\rangle=\frac{{\cal N}}{\lambda^r}\,
	\frac{\prod_{j=1}^m\Gamma(b_j+r\beta_j)\,
              \prod_{j=1}^n\Gamma(1-a_j-r\alpha_j)}
	     {\prod_{j=m+1}^q\Gamma(1-b_j-r\beta_j)\,
	      \prod_{j=n+1}^p\Gamma(a_j+r\alpha_j)}.
	\eqno(\mbox{A.11})
$$
Many of the classical probability densities are special cases of the
${\sf H}$-function distribution and can be written in the form of (A.9).
This was utilized in Eq.~(3.4) which enabled us to express the
Poisson transform of the scaling function~(2.4) analitically as
a discrete ${\sf H}$-function distribution.

\newpage\null\noindent
{\LARGE\bf References}
\vskip.5cm

\begin{tabbing}
 [00] \=  akarmiakarmiakarmiakarmiakarmiakarmiakarmiakarmiakarmi	\kill
 [1]  \>  I.M. Dremin, {\it Physics Uspekhi\/} 37 (1994) 715.       	\\
      \>  Yu.L. Dokshitzer, V.A. Khoze and S.I. Troyan, in
          {\it Perturbative QCD\/}, 					\\
      \>  ed. A.H. Mueller, World Scientific, 1989. 			\\
 {[}2]\>  Z. Koba, H.B. Nielsen and P. Olesen,
          {\it Nucl. Phys. B\/} 40 (1972) 314.                       	\\
 {[}3]\>  F. Cuypers and K. Tesima, {\it Z. Phys. C\/} 54 (1992) 87. 	\\
 {[}4]\>  Yu.L. Dokshitzer, {\it Phys. Lett. B\/} 305 (1993) 295. 	\\
 {[}5]\>  I.M. Dremin, {\it Mod. Phys. Lett. A\/} 8 (1993) 2747.	\\
 {[}6]\>  I.M. Dremin and V.A. Nechitailo,
          {\it JETP Lett.\/} 58 (1993) 881.				\\
 {[}7]\>  SLD Collab., K. Abe et al.,
	  {\it Phys. Lett. B\/} 371 (1996) 149. 			\\
 {[}8]\>  A.M. Polyakov, {\it Zh. Eksp. Teor. Fiz.\/} 59 (1970) 542.	\\
 {[}9]\>  N.G. Antoniou et al., {\it Phys Rev. D\/} 14 (1976) 3578.     \\
      \>  S. Krasznovszky and I. Wagner,
	  {\it Nuovo Cim. A\/} 76 (1983) 539.    			\\
      \>  P. Carruthers, {\it LA-UR-84-1009\/}, 1984.			\\
      \>  S. Krasznovszky and I. Wagner,
          {\it Phys. Lett. B\/} 306 (1993) 403. 			\\
      \>  R. Botet and M. P\l oszajczak, {\it GANIL-P-96-07\/}, 1996.   \\
 {[}10]\> E. Jahnke and F. Emde, {\it Tables of Higher
	  Functions\/}, Teubner, 1966.					\\
 {[}11]\> N.L. Johnson and S. Kotz, {\it Distributions in Statistics\/},
          {\it Vol. 2.,}					\\
       \> {\it Continuous Univariate Distributions\/},
	  Wiley, 1970.					\\
 {[}12]\> L. Amoroso, {\it Ann. Mat. Pura Appl.\/} 21 (1925) 123.       \\
       \> P. Rosin et al.,
	  {\it Bericht C 52 des Reichkohlenrates\/}, Berlin, 1933. 	\\
       \> S. Nukiyama and Y. Tanasawa,
	  {\it Trans. Soc. Mech. Japan\/} 5 (1939) 62.             \\
       \> E.W. Stacy, {\it Ann. Math. Statist.\/} 33 (1962) 1187.	\\
 {[}13]\> D.S. McKenzie, {\it Phys. Reports\/} 27 (1976) 37.            \\
 {[}14]\> A. Schenzle and H. Brand,
          {\it Phys. Rev. A\/} 20 (1979) 1628.				\\
       \> K. Lindenberg and V. Seshadri,
          {\it J. Chem. Phys.\/} 71 (1979) 4075.			\\
 {[}15]\> A.I. Golokhvastov, {\it Sov. J. Nucl. Phys.\/}
	  27 (1978) 430.						\\
 {[}16]\> R. Szwed, G. Wrochna and A.K. Wrob\l ewski,                   \\
       \> {\it Mod. Phys. Lett. A\/} 5 (1990) 1851
	  and 6 (1991) 245.						\\
 {[}17]\> P. Carruthers and C.C. Shih,
          {\it Int. J. Mod. Phys. A\/} 2 (1987) 1447.			\\
 {[}18]\> S.J. Orfanidis and V. Rittenberg,
	  {\it Phys. Rev. D\/} 10 (1974) 2892.				\\
 {[}19]\> B.D. Carter and M.D. Springer,
	  {\it SIAM J. Appl. Math.\/} 33 (1977) 542.			\\
 {[}20]\> A.M. Mathai and R.K. Saxena,
	  {\it The H-Function with Applications in}			\\
       \> {\it Statistics and Other Disciplines\/}, Wiley Eastern, 1978.\\
 {[}21]\> F. Oberhettinger,
	  {\it Tables of Fourier Transforms and Fourier}		\\
       \> {\it Transforms of Distributions\/}, Springer-Verlag, 1990.   \\
 {[}22]\> S. Krasznovszky,
	  {\it Mod. Phys. Lett. A\/} 8 (1993) 483.			\\
 {[}23]\> L. Bondesson, {\it Ann. Probab.\/} 7 (1979) 965.		\\
 {[}24]\> Aleph Collab., D. Buskulic et al.,
	  {\it Z. Phys. C\/} 69 (1995) 15.				\\
       \> Delphi Collab., P. Abreau et al.,
	  {\it Z. Phys. C\/} 50 (1991) 185.				\\
       \> L3 Collab., B. Adeva et al.,
	  {\it Z. Phys. C\/} 55 (1992) 39.				\\
       \> Opal Collab., P.D. Acton et al.,
	  {\it Z. Phys. C\/} 53 (1992) 539.				\\
       \> SLD Collab., J. Zhou, {\it private communication\/}, 1996.    \\
 {[}25]\> R.L. Prentice, {\it Biometrika\/}, 61 (1974) 539.		\\
       \> V.T. Farewell and R.L. Prentice,
	  {\it Technometrics\/}, 19 (1977) 39.			\\
 {[}26]\> R. Ugoccioni, A. Giovannini and S. Lupia,
	   {\it Phys. Lett. B\/} 342 (1995) 387.
\end{tabbing}

\vspace{2.5cm}
\noindent
FIGURE CAPTIONS

\vskip1cm
\noindent
Fig. 1a:\ \ Sign-changing oscillations of $H_q=K_q/F_q$ with $F_q$
given by Eq.~(3.10) for $k=3/2$. The neighbouring bumps at a fixed~$q$
are $H_q$-intervals of opposite sign in $\mu$.
For $\mu\leq1$ $H_q$ is always positive. The $\mu$-scale is
logarithmic and for clarity only the odd-rank moment ratios are shown.

\vskip.5cm
\noindent
Fig. 1b:\ \ Slices through Fig. 1a with $q=5$, 20 (left) and with
$\mu=1$, 3/2 (right).

\vskip.5cm
\noindent
Fig. 2a-b:\ \ The best fit to the SLD data for
$P_n$ and $H_q$ as described in the text. The fit parameters and the
quality of fits are shown in the last two rows of Table 2.

\newpage
\pagestyle{empty}

\begin{tabular}{||l|c|c|c||} \hline
{\sc Distribution}& \ shape par. \ & \ scale par.\ & \ exponent \ \\
 \hline\hline
 Generalized gamma      &   $k$     &   $\lambda$  &  $\mu$   \\ \hline\hline
 Gamma                  &   $k$     &   $\lambda$  &    1     \\
 Chi-square, $n$ d.o.f. &  $n/2$    &      1/2     &          \\
 Exponential            &    1      &   $\lambda$  &          \\ \hline\hline
 Weibull                &    1      &   $\lambda$  &  $\mu$   \\ \hline\hline
 Stratonovich           &   $k$     &   $\lambda$  &    2     \\
 Rayleigh               &    1      &              &          \\
 Maxwell molecular speed&   3/2     &              &          \\
 Maxwell mol. velocity  &   1/2     &              &          \\ \cline{1-3}
 Chi, $n$ d.o.f.        &  $n/2$    &  $1/\sqrt 2$ &          \\
 Half-normal            &   1/2     &              &          \\
 Circular normal        &    1      &              &          \\
 Spherical normal       &   3/2     &              &          \\ \hline\hline
 Pearson type V         &   $k$     &   $\lambda$  &  $-1$    \\
 Asymmetric L\'evy,
 $\alpha=1/2$           &   1/2     &              &          \\ \hline\hline
 Lognormal              &$k\to\infty$&  $\lambda$  &$\mu\to0$ \\ \hline\hline
 Pareto               & $k\to0$& $\lambda$ & $\mu\to\pm\infty$\\ \hline
\end{tabular}
\vspace{.5cm}
\begin{center}
Table 1. \ Special- and limiting cases of the generalized gamma density~(2.4)
\end{center}

\vspace{1.5cm}

\begin{tabular}{||l|c|c|c|c||} \hline
{\sc Experiment} & $\chi^2/\mbox{d.o.f.}$  &   $p$
                 & $\langle n\rangle$      &  $\sigma$    \\ \hline\hline
     Aleph       & \ 4.2/24&  $-0.230\pm0.311$  &  $20.970\pm0.407$
                 & $0.200\pm0.014$  \\
     Delphi      & 30.2/23 &  $-0.089\pm0.113$  &  $21.311\pm0.140$
                 & $0.199\pm0.006$  \\
     L3          & 14.9/20 &  $-0.123\pm0.238$  &  $20.660\pm0.349$
                 & $0.205\pm0.016$  \\
     Opal        & 10.7/24 &  $-0.124\pm0.130$  &  $21.370\pm0.128$
                 & $0.208\pm0.008$  \\
     SLD         & 31.2/20 &  $-0.001\pm0.060$  &  $20.892\pm0.068$
                 & $0.204\pm0.003$  \\ \hline\hline
     SLD ($H_q$) & 16.6/13 &  $\ \ 0.060\pm0.029$  &  $21.346\pm0.126$
                 & $0.214\pm0.004$  \\ \hline
\end{tabular}

\vspace{1cm}\noindent
\hskip0cm Table 2. \ The results of fits to the $e^+e^-$ multiplicity data
at the $Z^0$ peak

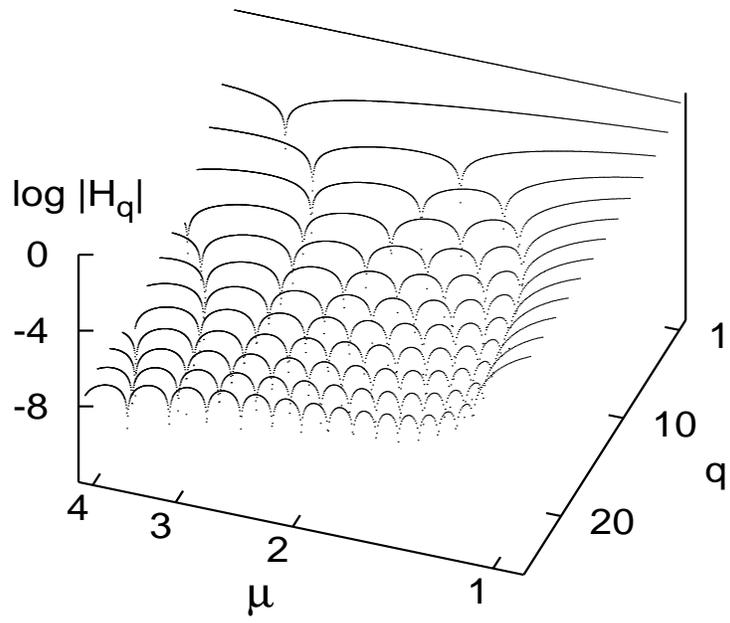
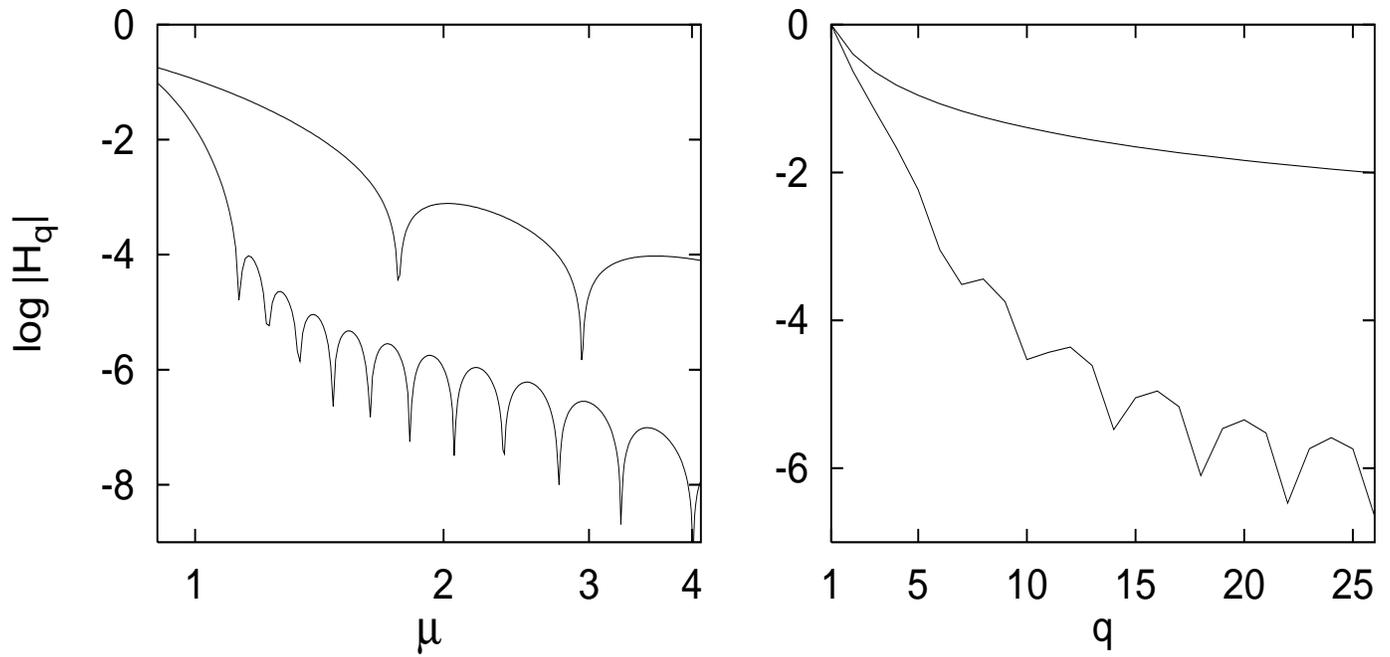
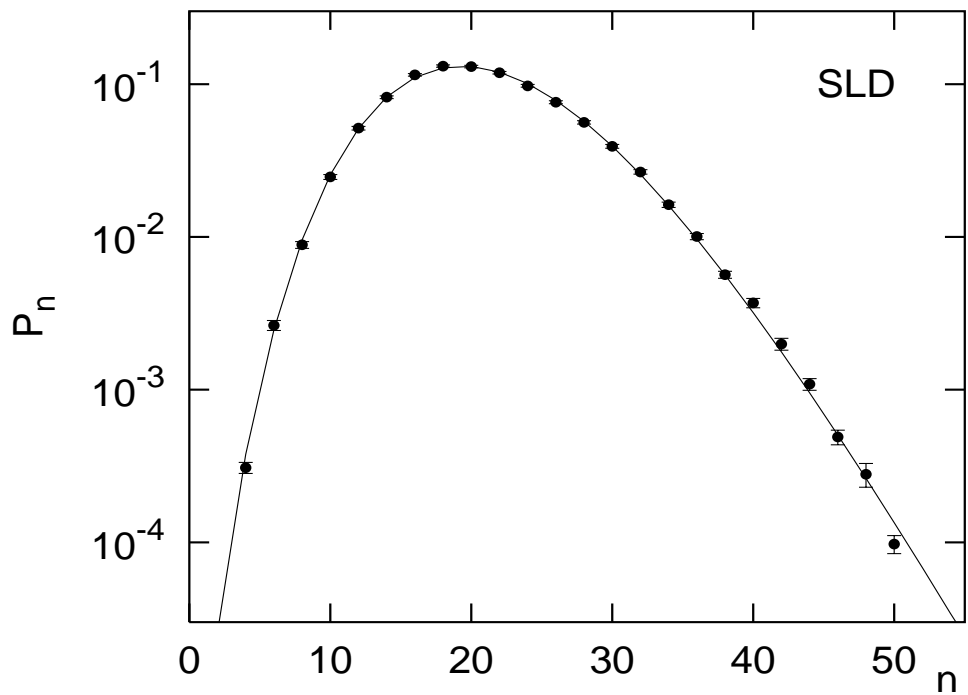
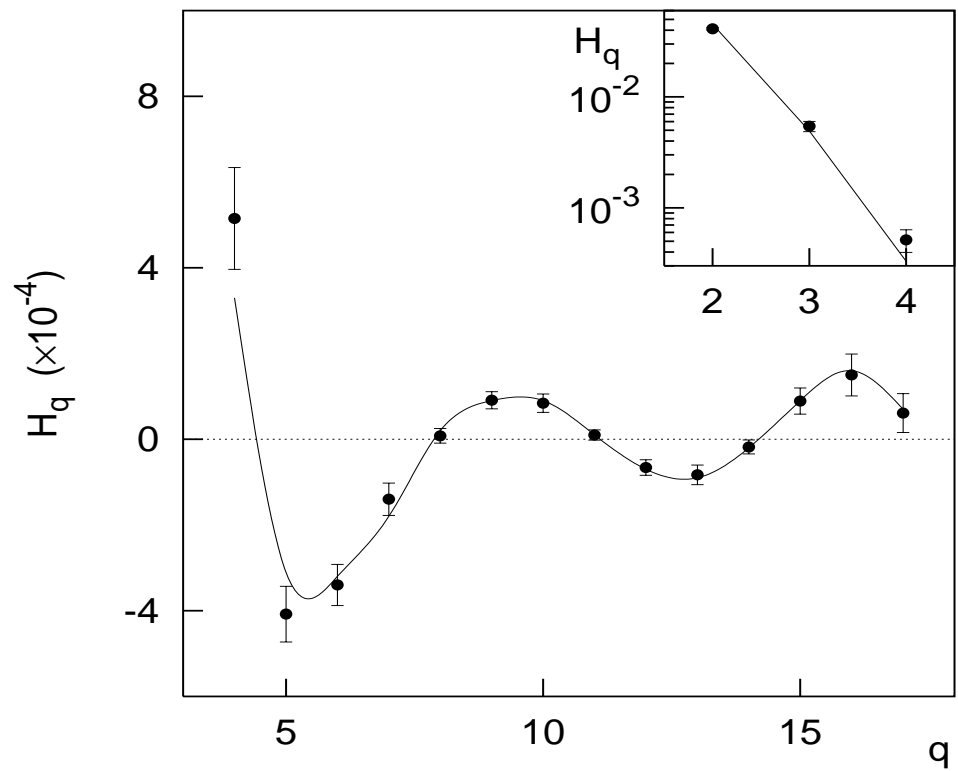
\end{document}